# Thermal Raman study of $Li_4Ti_5O_{12}$ and discussion about the number of its characteristic bands

Aleksey A. Nikiforov [1], Alexander S. Krylov [2], Svetlana N. Krylova [2], Vadim S. Gorshkov [3] and Dmitry V. Pelegov [1,*]

[1] Institute of Natural Sciences and Mathematics, Ural Federal University, Ekaterinburg, 620002, Russia

[2] L.V. Kirensky Institute of Physics SB RAS, Krasnoyarsk, 660036, Russia

[3] Independent researcher, Ekaterinburg, 620142, Russia

* Corresponding author: dmitry.pelegov@urfu.ru

**Abstract**

Lithium battery industry is booming, and this fast growth should be supported by developing industry-friendly tools to control the quality of positive and negative electrode materials. Raman spectroscopy was shown to be a cost-effective and sensitive instrument to study defects and heterogeneities in lithium titanate, popular negative electrode material for high-power applications, but there are still some points to be clarified. This work presents a detailed thermal Raman study for lithium titanate and discusses the difference of the number of predicted and experimentally observed Raman-active bands. The low-temperature study and the analysis of thermal shifts of bands' positions during heating let us to conclude about advantages of the proposed approach with surplus bands and recommend using shifts of major $F_{2g}$ band to estimate the sample heating.

**Key words:**

Raman spectroscopy, lithium titanate, lithium titanium oxide, spinel structure, defects, distortions

## 1. Introduction

Lithium titanate ($Li_4Ti_5O_{12}$, LTO) with a spinel-like structure is an efficient commercial anode material used in lithium batteries. The numerous advantages of natural and artificial graphite turn LTO into a niche product, but its market share remains rather stable. The main applications of LTO-based lithium cells are high-power traction batteries for some electric buses and long-lasting stationary power storage systems.

Raman spectroscopy is a popular tool for structural characterization of lithium battery materials and LTO is among the most studied ones. The very first paper with Raman spectra of LTO was published in 1983 [1], but whereas it was published prior to lithium battery boom, has no electronic version and was written in Russian, currently it is out of the view of researchers. The next two articles were published only two decades later in 2003 [2,3]. The detailed one, written by Leonidov et al [2], was also published in Russian journal but since it had an English version, it became much more popular and currently is commonly considered as the first publication with Raman spectra of LTO.

The first thermal study of Raman spectra evolution was presented in the same pioneering work of Leonidov et al [2]. Twelve years later, in 2015, it was followed by works of Knyazev et al [4] and Mukai et al [5]. It may seem that this topic is closed for LTO, but we believe that some of the results presented in these papers can be revised. The point is that in the work [2] the analysis of band position evolution with heating has revealed an anomalous shift to the larger values for $E_g$ and $A_{1g}$ bands, while others shifted more predictably to lower values. In the work [5] $A_{1g}$ band was alternatively reported to shift to lower values with heating, while the results for $E_g$ band were not reported.

The ambiguity of Raman band shifts with heating can be caused by the uncertainty of the exact number of experimentally observed vibrational modes for LTO. Symmetry analysis usually predicts five Raman-active vibrational modes, namely $A_{1g} + E_g + 3F_{2g}$ but all the published Raman spectra of LTO contains larger number of characteristic bands. In the work [6] we have published the statistics of Raman band positions, which have demonstrated that the surplus bands are discussed rarely but regularly. In the same work the surplus band near 520 cm$^{-1}$ was attributed to the presence of defects and further the idea was elaborated in the work [7] where other surplus characteristic bands of LTO were also discussed in terms of imperfections. The presence of surplus bands, not compatible with the selection rules, is not LTO-specific feature and was also discussed for other materials with spinel structure [8–10].

In this work we will focus on the surplus bands of $Li_4Ti_5O_{12}$ with a spinel structure and use thermal Raman study to validate its amount, required for more accurate Raman spectra deconvolution. Also, we will calculate thermal shifts of main characteristic bands to use it further for estimation of possible sample heating during Raman spectra measurements.

## 2. Materials and methods

LTO samples were prepared by the usual solid-state synthesis method using lithium carbonate ($Li_2CO_3$) and titanium dioxide ($TiO_2$) in anatase form. During slow heating in the temperature range of 500–700°C lithium carbonate decomposes into carbon dioxide and lithium oxide ($Li_2O$), which, in its turn, reacts with titanium oxide to set up lithium titanate $Li_4Ti_5O_{12}$ and other intermediate substances of $Li_2O – TiO_2$ system. Synthesis was carried out by four stages, during which the desired product was fully formed, which was confirmed by X-Ray diffraction (XRD) analysis. More information about sample synthesis can be found in the works [11,12].

Raman spectra were measured using two different experimental set-ups. Integral measurements of the compacted LTO powder in the temperature range 8-345 K were performed in a dynamic mode using a triple spectrometer T64000 (Horiba Jobin Yvon) in subtraction dispersion mode. Spectra excitation was an DPSS laser Spectra-Physics Excelsior-532-300-CDRH at of 532 nm. Signal accumulation time was 30 s; spectral resolution was 2 cm$^{-1}$; CCD matrix pixel coverage was about 0.3 cm$^{-1}$, ramping was 0.8 K/min; temperature step was 0.8 K; laser excitation power on a sample was 5 mW. The dynamic temperature technique is described in detail in the work [13].

The temperature dependences of the Raman scattering spectra for individual particles deposited on a glass substrate were measured using a confocal Raman microscope Alpha 300 RA (WITec GmbH) equipped with a solid-state laser with a wavelength of 488 nm. The laser beam was focused by a lens with a large working distance of 100× (NA 0.75). The scattered light was collected by the same lens in the backscattering geometry and passed through an edge filter. An optical fiber with a diameter of 50 microns was used as a confocal pinhole. A diffraction grating with 1800 grids per mm was used to decompose the scattered light into a spectrum, which was then recorded by a CCD camera. The deconvolution of the experimentally obtained Raman spectra was made using "Wolfram Mathematica" with 6 (5+1) and 12 (5+7) peaks.

Thermal measurement of individual particles was investigated using a temperature-controlled stage THMS600 and a system controller T96 (Linkam Sci. Inst.). The sample was heated in the range from 25 °C to 400 °C in air with a constant heating rate of 10°C/min. After reaching a certain point value, the temperature remained constant, and the Raman spectra were measured only after 10 minutes required for uniform heating of the system. The laser power was set to 8.6 mW, measuring the power directly in the output beam through the viewing window of the temperature-controlled stage using a Vega laser power and energy meter with a PD300R-3W photodiode laser sensor (Ophir Photonics).

The density functional theory (DFT) calculations were carried out by the plane-wave pseudo-potential method using Cambridge Serial Total Energy Package code (CASTEP) [14]. The calculations were performed for Li: 2s1; O: 2s2, 2p4; Ti: 3s2, 3p6, 3d2, 4s2. The structures were relaxed using the Broyden, Fletcher, Goldfarb and Shannon (BFGS) minimization method algorithm [15]. The lattice constants and atom coordinates were optimized by minimizing the total energy. Through a series of convergence studies concerning cut-off energies and k-points, the cut-off energies were set to 860 eV, and the K-space integration over the Brillouin zone was carried out using a 2×2×1 k-point Monkhorst-Pack mesh [16]. All these parameters were tested to ensure that the self-consistent total energies converged to within 1.0 x 10$^{-9}$ eV/atom. We have made the structural relaxation and optimization with the approximations, to determine the internal atomic coordinates and structure parameters. The PBEsol (GGA) functional has been used [17].

## 3. Results and Discussion

## 3.1. Integral low-temperature measurements

The Raman spectra of LTO consist of a set of relatively broad overlapping bands [2,18,19]. The most obvious way to facilitate the deconvolution of Raman spectra is to measure them at low temperatures.

The first series of Raman spectra was recorded in the temperature range from 8 to 345 K (from −265 to 72 °C) for compacted powder sample (Fig. 1, Fig. S1). Even the first look at the LTO Raman spectrum at low temperatures clearly demonstrates that the standard "5+1" approach with a set of 5 predicted bands with one $A_{1g}$ satellite is not enough. The "surplus" unidentified bands can be divided into two categories, let us denote them as "separate" and "satellite" peaks.

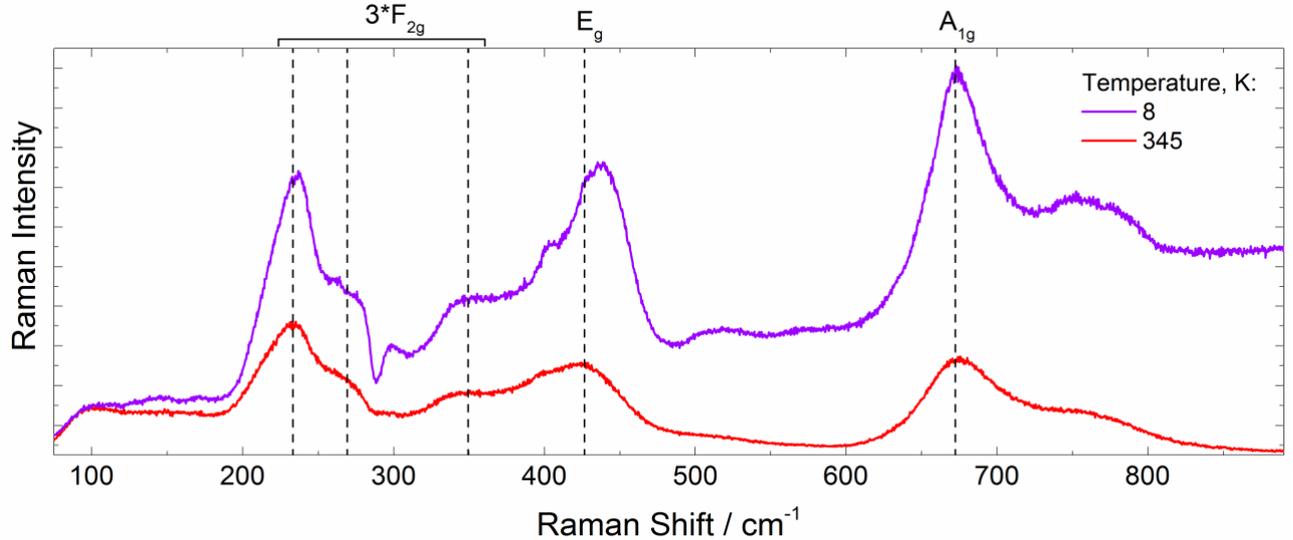

Fig. 1. Raman spectra of LTO at 8 and 345 K. The Raman spectra at intermediate temperatures are given at Fig. S1.

### 3.1.1. Separate "surplus" bands

The most frequently mentioned separate "surplus" band is a broad peak at about 520 cm$^{-1}$, hereafter denoted as $X_{520}$. This peak is observed in almost all publications with LTO Raman spectra and is usually attributed to the presence of defects [6,7,20]. Alternatively, some authors attribute it to the presence of near phases, but the work [21], which studied incomplete synthesis with both $Li_2TiO_3$ and rutile $TiO_2$ had shown that both phases have a clear wide "valley" at the location of $X_{520}$ band. In the work [6] the "defect" origin of this band was indirectly confirmed by the experiment. For our sample, $X_{520}$ peak broadens with the temperature growth and becomes less pronounced at the room temperature, indirectly confirming the high quality of the studied sample.

A peak at 310 cm$^{-1}$ (hereafter denoted as $X_{310}$) is another separate "surplus" band clearly seen at low temperatures [4,5], but at the room temperature and above it becomes almost indistinguishable. Contrary to $X_{520}$ band it is usually not mentioned in literature but can be somehow supposed in many published Raman spectra.

A plateau below 200 cm$^{-1}$ let us suppose a superposition of several bands, but the absence of clearly resolved peaks makes the analysis of this spectral region obstructed in our case. Even though the study of "surplus" bands below 200 cm$^{-1}$ is rather promising due to the fact of variation of bands number and position from paper to paper, in this work we will exclude it from our consideration.

Besides the mentioned above separate "surplus" bands the Raman spectra of LTO have satellite "surplus" bands, located close to main bands. Such satellite bands are often hidden and overlapped by major band shoulder but can drastically influence the Raman spectra parameter of their neighbors. In order to illustrate this, let us consider three main spectral regions: 180-330 cm$^{-1}$, 300-500 cm$^{-1}$, and 630-840 cm$^{-1}$ (Fig. 2).

### 3.1.2. Satellite "surplus" bands in the spectral range 180-330 cm$^{-1}$

The standard approach implies the presence of only two theoretically predicted $F_{2g}$ bands – major at about 230 cm$^{-1}$ and minor at about 275 cm$^{-1}$. Two bands work rather well at room temperatures (Fig. 2c), but not at −265 °C (Fig. 2a). The first problem is that a fitting software usually tries to broaden and shift the minor $F_{2g}$ band to somehow compensate the absence the $X_{300}$ band. The second problem is that a gaze at the spectra in this range let us to suppose the presence of at least one more "surplus" band (hereafter denoted as $X_{260}$), hidden in-between two $F_{2g}$ bands (Fig. 2b). The steep right shoulder of the minor $F_{2g}$ band indicates its narrowness and supposes a valley between major and minor $F_{2g}$ bands. But instead of a valley one can see a small but clear peak, thus proving the need to use the $X_{260}$ band.

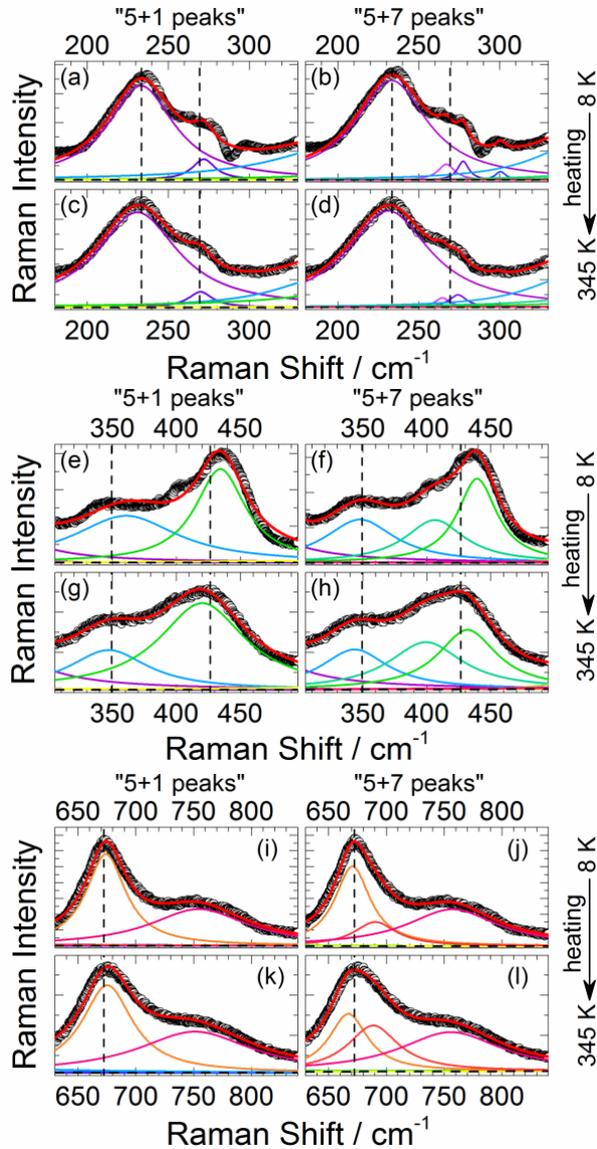

Fig. 2. The comparison of (a),(c),(e),(g),(i),(k) standard and (b),(d),(f),(h),(j),(l) alternative approaches for deconvolution of LTO Raman spectra at temperatures (a),(b),(e),(f),(i),(j) 8 and (c),(d),(g),(h),(k),(l) 345 K in three spectral ranges: (a)-(d) 180-330 cm$^{-1}$, (e)-(h) 300-500 cm$^{-1}$, and (i)-(l) 630-840 cm$^{-1}$.

Besides the presence of $X_{260}$ and $X_{300}$ band, clearly seen at low temperatures, we want to mention a possible asymmetry of major $F_{2g}$ band or one more small "surplus" band at about 200 cm$^{-1}$. For the studied sample, the using of possible $X_{210}$ band is not reasonable, but in the work [7] the presence of $X_{200}$ band looked rather realistic. So even though we cannot prove its existence we propose to carefully study the LTO Raman spectra below 220 cm$^{-1}$ in future.

### 3.1.3.  Satellite "surplus" bands in the spectral range 300-500 cm$^{-1}$

While the presence of satellite $X_{260}$ band became an unexpected result of our low temperature study, the satellite $X_{400}$ band at about 400 cm$^{-1}$ was the main objects of our research. At the room temperature, the two-peak fitting in this spectral range looks if not perfect but acceptable (Fig. 2g). But at low temperatures, two-peak fitting becomes unsatisfactory (Fig. 2e). Using excess $X_{400}$ band greatly helps and makes room temperature fitting near perfect and low temperature fitting acceptable.

In fact, $X_{400}$ band alone does help but not fully. Adding one more satellite band between $F_{2g}$ and $X_{400}$ bands will obviously increase the fitting quality (Fig. 1, 2f), but in this work we will not do so. The main idea of our work is not to provide the best fitting but to study how additional bands will the change of temperature dependence of five theoretically predicted bands.

### 3.1.4. Satellite "surplus" bands in the spectral range 630-840 cm$^{-1}$

The presence of an unknown satellite to the right of $A_{1g}$ is so obvious that it cannot be ignored (Fig. 2i-l). The question is, is there only one satellite? For example, Raman spectrum of polycrystalline $NiAl_2O_4$ with a spinel structure contains more than one excess band [8], so why not to suppose more $A_{1g}$ satellites for LTO with spinel structure?

Adding two satellites ($X_{690}$ and $X_{750}$) instead of one have gave us ambiguous result. From one hand, this makes fitting better (compare Fig. 2i,k with Fig. 2jl). From the other hand, the improvement is minor, and the additional "surplus" band greatly decreases the intensity of major $A_{1g}$ band. To understand which fitting option to choose, let us continue with Raman spectra measurements above room temperatures and compare our alternative "5+7" fitting with the standard "5+1" one.

### 3.2. Measurements of single LTO particles above room temperature

Raman spectra at the 25, 100, 200, 250, 300, 350, and 400 °C were measured for 10 single submicron particles (or small aggregates). Using single-particle approach help us to analyze the variation Raman spectra parameters and later average them for better statistics. Also, such an approach let us to conclude that contrary to Mn-doped LTO [22], all 10 particles of pristine LTO maintain characteristic Raman spectra under the combined action of heat and probing laser radiation. Some changes, in fact, do observed and discussed in the Supplementary Materials (Fig. S4) but these alterations are minor.

In the work [2] the authors laid down a standard approach using 5+1 peaks. They have concluded that the heating results in $F_{2g}$ and $E_g$ bands move towards each other (black symbols and lines at Fig. 3), and both changes were deviant. $F_{2g}$ band abnormally moves to the higher wavenumbers, which is more specific for cooling and don't comply with thermal behavior of the two other $F_{2g}$ bands. $E_g$ band with heating normally moves toward lower wavenumbers, but the value of this temperature shift is abnormally high comparing with all other bands (Fig. 3). Thermal shift of $A_{1g}$ band toward higher wavenumbers is also abnormal in this work [2].

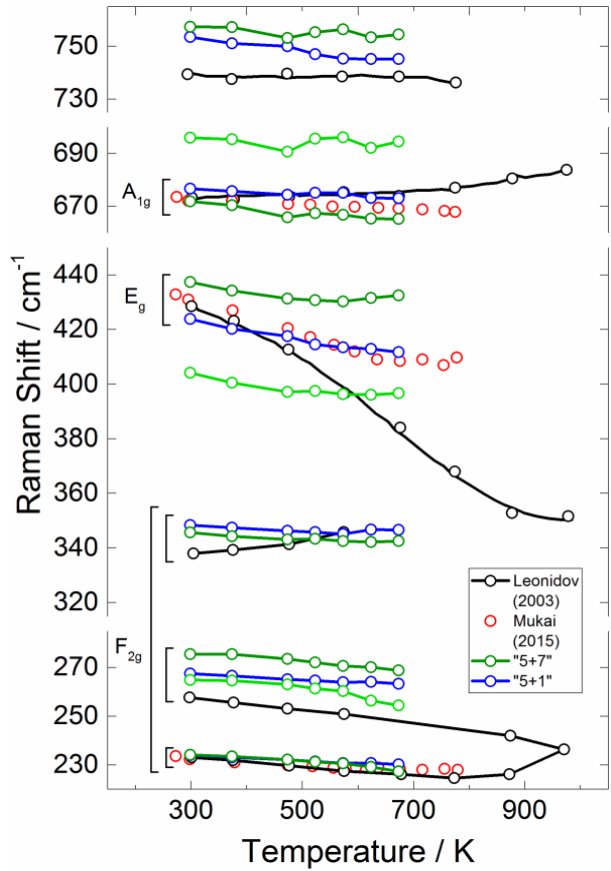

Fig. 3. Shifts of the band positions during heating. Our results (green and blue symbols and lines) are compared with ones, reconstructed from the work of Leonidov et al [2] (black symbols and lines) and Mukai et al [5] (red symbols). Blue corresponds to standard approach with 5+1 peaks. Green corresponds to the proposed alternative approach with 5+7 peaks.

The results, presented in the work [5], differ and thermal evolutions of $A_{1g}$ and $E_g$ bands are rather normal (Red circles at Fig. 3). The only problem is the absence of data for two more $F_{2g}$ bands. Please mention that the normal thermal shift toward lower wavenumbers were reported and for other electrode materials for lithium batteries, for example LFP [23] and LCO [24].

In our case, the fitting using 5+1 peaks almost comply with results, shown in work [5] and demonstrates normal thermal shifts for all characteristic bands (blue symbols and lines at Fig. 3). Using "5+7" peaks approach, at first glance, don't give a strong difference, comparing with "5+1" peak approach (compare blue and green symbols at Fig. 3), but this is a misleading impression and a more detailed look at these dependences reveals a notable difference (Fig. S2 and S3, Table 1). Table 1 contains band positions and FWHM for five main predicted bands at 25 and 400 °C, which were obtained using "5+1" peaks and "5+7" peaks approaches. Also, there are corresponding thermal shifts (dν/dT) and thermal broadenings (dω/dT) for these bands, as well as the differences Δ of values between "5+1" peaks and "5+7" peaks fitting at 25 and 400 °C.

Table 1. Band positions and FWHMs for five main characteristic LTO bands, obtained using two alternative approaches. dν/dT and dω/dT are a thermal shift and thermal broadening of the corresponding bands. Δ(T) is the difference between two approaches at the given temperature T.

|  |  | $F_{2g}$ | $F_{2g}$ | $F_{2g}$ | $E_g$ | $A_{1g}$ |
|---|---|---|---|---|---|---|
|  |  | Band position ν, cm$^{-1}$ | | | | |
| 5+1 bands | 25 °C | 234.1 | 267.6 | 348.4 | 423.9 | 676.6 |
|  | 400 °C | 230.2 | 263.4 | 346.6 | 411.7 | 673.0 |
|  | dν/dT, cm$^{-1}$/100°C | −1.2 | −1.3 | −1.2 | −3.8 | −0.6 |
| 5+7 | 25 °C | 234.1 | 275.6 | 345.7 | 437.4 | 671.8 |

|   |   | 400 °C | 227.3 | 268.9 | 342.5 | 432.5 | 665.2 |
|---|---|---|---|---|---|---|---|
|   |   | dν/dT, cm$^{-1}$/100°C | −1.3 | −1.9 | −1.1 | −2.6 | −2.0 |
|   | Δ(25 °C) |   | 0 | 8.0 | −2.7 | 13.5 | −4.8 |
|   | Δ(400 °C) |   | −2.9 | 5.5 | −4.1 | 20.8 | −7.8 |
|   |   | Band FWHM ω, cm$^{-1}$ | | | | | |
| 5+1 bands | 25 °C |   | 48.6 | 23.0 | 68.8 | 88.3 | 59.9 |
|   | 400 °C |   | 73.2 | 18.5 | 78.1 | 99.0 | 70.8 |
|   | dω/dT, cm$^{-1}$/100°C |   | 7.3 | −0.7 | 3.1 | 4.0 | 3.0 |
| 5+7 bands | 25 °C |   | 56.9 | 13.3 | 73.2 | 57.6 | 51.1 |
|   | 400 °C |   | 78.5 | 15.4 | 77.2 | 78.5 | 58.8 |
|   | dω/dT, cm$^{-1}$/100°C |   | 6.0 | 2.7 | 1.2 | 7.2 | 2.4 |
|   | Δ(25 °C) |   | 8.3 | −9.7 | 4.4 | −30.7 | −8.8 |
|   | Δ(400 °C) |   | 5.3 | −3.1 | −0.9 | −20.5 | −12 |

Consider, for example, the major $F_{2g}$ band at about 230 cm$^{-1}$ (first column at Table 1). Illustratively, for this band Δ(25 °C)=0, which means that at the room temperature its position is the same for both "5+1" and "5+7" approaches (Table 1). But at the 400 °C this difference reaches almost 3 cm$^{-1}$. We suppose that since heating decreases signal/noise ratio for Raman spectra, the minor $F_{2g}$ band's shoulder becomes less pronounced and fitting software tries to fit both $F_{2g}$ bands mainly by major $F_{2g}$ band while the minor $F_{2g}$ band shifts toward $X_{300}$ band to compensate its broadening.

Please mention that "5+7" approach provide lower dispersion of thermal shift values: from 1.1 to 2.6 cm$^{-1}$/100°C, while for "5+1" approach thermal shifts values range from 0.6 to 3.8 cm$^{-1}$/100°C (Table 1). The larger dispersion of thermal shifts for "5+1" approach can indicate the lower fitting accuracy.

Table 1 can be used to estimate the temperature change by the shift of band positions. Comparing the two most intensive characteristic LTO bands, the major $F_{2g}$ and $A_{1g}$, we would recommend using the first one, since the major $F_{2g}$ band can boast of the lowest variation of band positions from particle to particle (Fig. S3a).

### 3.3. Interpretation of Raman spectra using alternative DFT calculation

All the results presented above led us to one clear finding – the standard five characteristic bands are not enough, and this fact can't be ignored. Let us try to illustrate the possible the origin of surplus bands of LTO using DFT calculations.

The key issue for factor group analysis is that it considers LTO as the perfect representative of space group $Fd\bar{3}m$, while Li[Li$_{1/6}$Ti$_{5/6}$]$_2$O$_4$ with some Li atoms on Ti positions has inherently distorted spinel-like structure. For the occupancy of all positions to be equal 1, it will be necessary to increase the size of the cell and lower the symmetry of the cell. With an increase in the cell size, the number of atoms entering it will increase, which means that the number of vibration modes will also increase.

The structure of Li$_4$Ti$_5$O$_{12}$ is distorted. We used ordered structure of Li$_4$Ti$_5$O$_{12}$ with lower symmetry in the DFT calculation. The crystallography information file from material project (mp685194) [26] have been used. The structure was ordered with using the algorithm for generating all derived superstructures of a non-primitive parent lattice [27]. The authors of the project used as the initial experimental data of the work [28]. In this theoretical material project Li$_4$Ti$_5$O$_{12}$ crystals belong to space group C2/c (No. 15), its point group is $C_{2h}$ (2/m). The experimental and calculated Raman spectra are presented in Figure 4, and it demonstrates that the theoretical spectrum agrees well with the experimental spectrum.

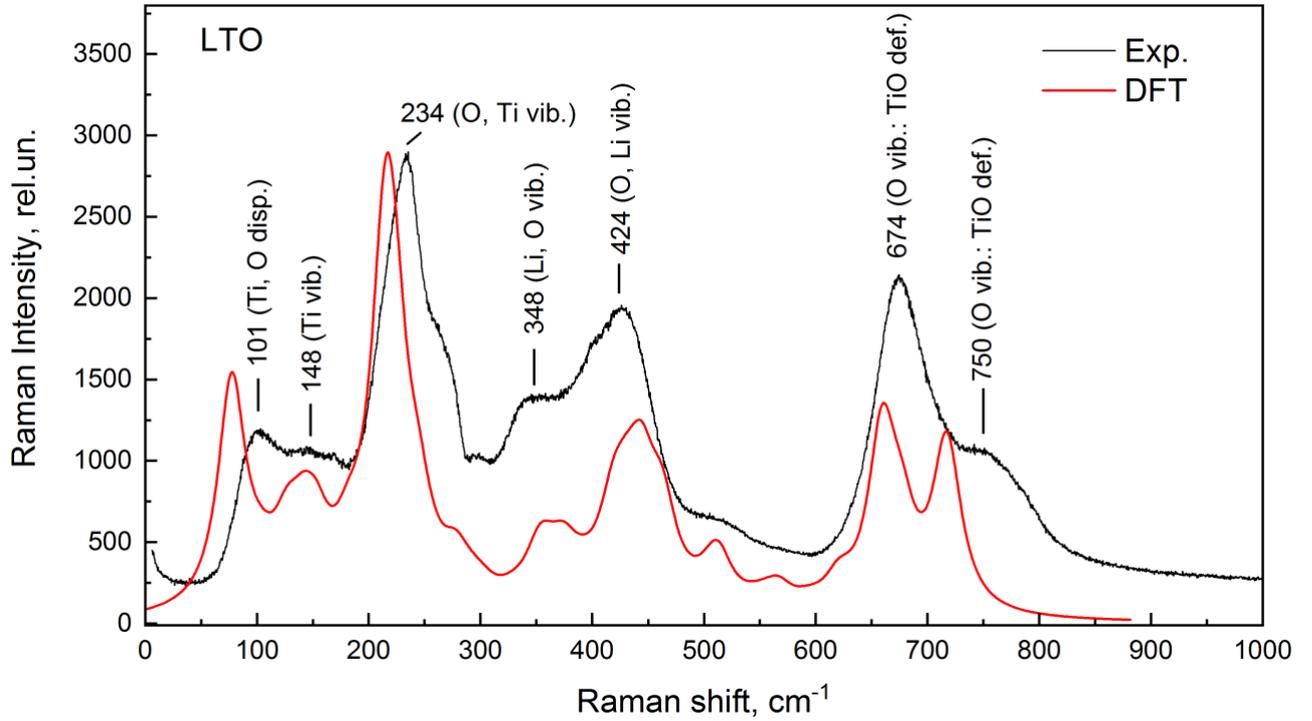

Fig. 4. The comparison of experimental and calculated Raman spectra of $Li_4Ti_5O_{12}$

Our experimental modes for LTO with a lower symmetry and LTO modes from other works [2,25,29], calculated modes and them assignments are presented in the Table 2. Strongest experimental mode of 233 cm$^{-1}$ has associated with Ti and O vibrations. It is full symmetry $A_g$ mode. High-frequency modes 750 and 674 cm$^{-1}$ associated with vibrations of oxygen atoms. The middle part of spectrum at 423 cm$^{-1}$ associated with vibrations of Li an O atoms. The mode at 101 cm$^{-1}$ includes Ti and O vibrations. The forms of the vibrations are presented in Table 2. The frequencies of all calculated vibrational modes are presented in the Table S3.

**Table 2.** Some experimental and calculated Raman modes of $Li_4Ti_5O_{12}$ with them assignments

| Experimental, cm$^{-1}$ | | Sym. $Fd\bar{3}m$ phase | Calc, cm$^{-1}$ C2/c phase, this work | Form of vibrations (C2/c phase) this work | Assignment (C2/c phase) this work |
|---|---|---|---|---|---|
| this work | other works | | this work | | |
| 750 | 743 [29], 745.5 [25], 740 [2] | $A_{1g}$ [29] (usually marked as a satellite) | 717 | 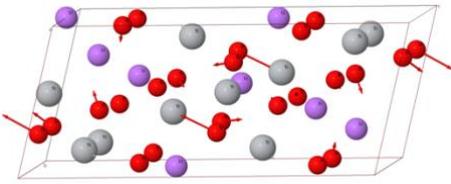 | O vibrations (TiO deformation) $A_g$ |
| 674 | 674 (680) [25], 673.7 [25], 675 [2] | $A_{1g}$ [29] | 659 | 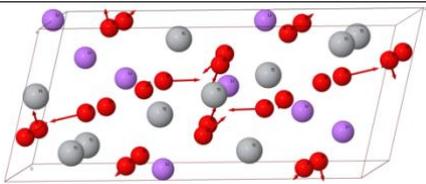 | O vibrations (TiO deformation) $A_g$ |
| 424 | 424 [29], 430.3 [25], 430 [2] | $E_{2g}$ [29] | 445 | 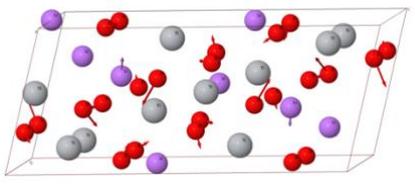 | O, Li vibrations $A_g$ |
| 348 | 340 [2], 340 [25], 344 [29] | $F_{2g}$ [29] | 352 | 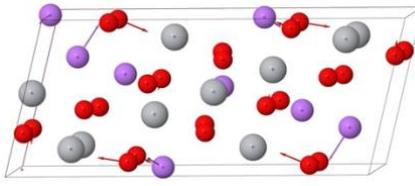 | Li, O vibrations $A_g$ |
| 234 | 233 [29], 234.6 [25], 235 [2] | $F_{2g}$ [29] | 214 | 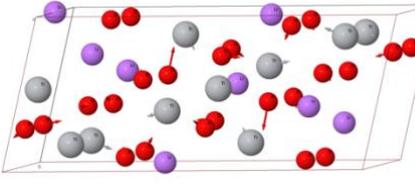 | O, Ti vibrations $A_g$ |
| 148 | 160 [25], 160 [2] | | 153 | 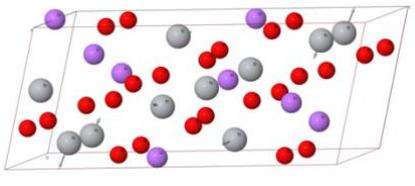 | Ti vibrations $A_g$ |
| 101 | 95.9 [25] | | 77 | 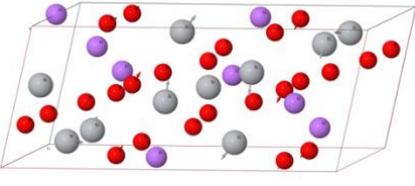 | Ti, O vibrations $B_g$ |

The DFT calculation based on material project (mp685194) [26] provides alternative interpretations of at least seven Raman modes. The highest frequency modes (600-800 cm$^{-1}$) are associated with the movement of oxygen atoms. Lithium and titanium atoms vibrate in the range of 300-500 cm$^{-1}$. The strongest vibration at 234 cm$^{-1}$ was interpreted as vibrations of oxygen and titanium atoms. The low-frequency vibrational mode 101 cm$^{-1}$ is also associated with vibrations of oxygen and titanium.

This alternative explanation is looking to be reliable up to date but this still an approximation. Anyway, this study shows that five Raman modes are not enough to describe an experimental Raman spectrum in LTO and an increase in the number of modes is because the LTO structure is inherently distorted.

## 4. Conclusions and methodological recommendations

The performed thermal study of pristine LTO led us to propose revising the standard "5+1 peaks" fitting approach and using an alternative "5+7 peaks" approach with additional excess bands. Some of these surplus bands are clearly seen even at room temperatures while the others were revealed only after the low-temperature measurements. The origin of these surplus bands is the subject of the further studies, but the need of its introduction was confirmed by the analysis of thermal shifts of characteristic bands' positions.

The shift of major $F_{2g}$ band at about 230 cm$^{-1}$ during heating or exposition to laser irradiation was proposed to be used to define a temperature change. In the case of "5+7" approach, the thermal shift is about $-1.3$ cm$^{-1}$/100°C. In the case of "5+1" approach, the thermal shift is a slightly lower and is about $-1.2$ cm$^{-1}$/100°C.

The performed DFT calculation, considering distorted LTO as belonging to space group C2/c (No. 15), shows good agreement with experiment, thereby indicating a new possible direction towards understanding the origin of the surplus bands.

**Funding**

The research was funded by the Russian Science Foundation (project No 22-22-00350, https://rscf.ru/project/22-22-00350).

**Acknowledgments**

The Raman spectra measurements were made using equipment of the Ural Center for Shared Use "Modern nanotechnology" Ural Federal University (Reg. No 2968), supported by the Ministry of Science and Higher Education of the Russian Federation (Project No 075-15-2021-677). Authors warmly thank Anastasia A. Koshkina for some samples.

**Conflicts of Interest:** The authors declare no conflict of interest.

**Data Availability Statement:** The datasets generated during and analyzed during the current study are available from the corresponding author on reasonable request.